\documentstyle[aps,prd,preprint]{revtex}
\begin{document}
\draft
\title{Conformally Coupled Induced Gravity with Gradient Torsion}
\author{Yongsung Yoon\footnote{E-mail: cem@hepth.hanyang.ac.kr}}
\address{Department of Physics, Hanyang University, Seoul 133-791, Korea}
\maketitle
\begin{abstract}
It is found that conformally coupled induced gravity with gradient torsion
gives a dilaton gravity in Riemann geometry. In the Einstein frame of the
dilaton gravity the conformal symmetry is hidden and a non-vanishing cosmological
constant is not plausible due to the constraint of the conformal coupling.
\end{abstract}
\section{Introduction}

Before the success of Weinberg-Salam model, the weak interaction was
characterized by the dimensional Fermi's coupling constant,
$G_{F}= (300 Gev)^{-2}$, far below the electro-weak scale.
But later it turns out that the dimensional coupling constant is the low
energy effective coupling which is determined by the dimensionless
electro-weak coupling constant and the vacuum expectation value of Higgs
scalar field through the spontaneous symmetry breaking.
The weakness of the weak interaction is originated from the large
vacuum expectation value of Higgs field \cite{Weinberg}.

From this lesson, it is suspected that gravity may be also
characterized by a dimensionless coupling constant $\xi$ with the gravitational
constant $G_{N}$ given by the inverse square of the vacuum
expectation value of a scalar field.
The weakness of the gravity can be associated with a symmetry breaking
at a very high energy scale.
It has been independently proposed by Zee \cite{Zee}, Smolin \cite{Smolin},
and Adler \cite{Adler} that the Einstein-Hilbert action can be replaced by
the induced gravity action
\begin{equation}
     S =  \int d^{4}x \sqrt{-g}\{\frac{1}{2}\xi \phi^{2} R +
          \frac{1}{2} \partial_{\mu}\phi \partial^{\mu}\phi - V(\phi)\},
\label{action}
\end{equation}
where the coupling constant $\xi$ is dimensionless.
The potential $V(\phi)$ is
assumed to have its minimum value at $\phi = \sigma$, then the
above action is reduced to the well known Einstein-Hilbert action
with gravitational constant $G_{N}=\frac{1}{8\pi\xi\sigma^{2}}$.

In the analogy of the $SU(2) \times U(1)$ symmetry of the
electro-weak interactions, we can consider a symmetry which may be broken through
a spontaneous symmetry breaking in the gravitational interactions.
The most attractive candidate symmetry is the Weyl's conformal symmetry which
rejects the Einstein-Hilbert action, but admits the induced gravity action
Eq.(\ref{action}) with a specific conformal coupling,
$\xi = \frac{1}{6}$.

In Riemann space, the conformal coupling is unique with $\xi=\frac{1}{6}$.
However, introducing the vector torsion, an extended conformal coupling is
possible in induced gravity \cite{yoon95a,yoon95b} because the vector torsion
plays the role of a conformal gauge field in Riemann-Cartan space \cite{Smolin,Nieh,yoon88}.

It is found that induced gravity at conformal coupling should have conformal
invariance for consistency \cite{yoon95b,yoon97}.
We investigate the conformal coupling in induced gravity with a gradient torsion.

\section{Conformal Couplings in Induced Gravity}

The induced gravity action Eq.(\ref{action}) is invariant under the
conformal transformation,
\begin{equation}
    g_{\mu\nu}'(x) = exp(2\rho) g_{\mu\nu}(x), ~~~
    \phi'(x) = exp(-\rho) \phi(x),
\end{equation}
at the conformal coupling $\xi = \frac{1}{6}$ for a conformally invariant
scalar potential.

In Riemann-Cartan space, an extension of conformal coupling with the torsion
in induced gravity is possible. It is found that the minimal extension to
Riemann-Cartan space is sufficient for our purpose.

The conformal transformation of the affine connections
$\Gamma^{\gamma}_{~\beta \alpha}$ is determined from the
invariance of the tetrad postulation,
\begin{equation}
       D_{\alpha} e^{i}_{\beta}\equiv\partial_{\alpha}e^{i}_{\beta} +
       \omega^{i}_{j\alpha} e^{j}_{\beta}-
       \Gamma^{\gamma}_{~\beta \alpha}e^{i}_{\gamma} = 0,
\label{cor}
\end{equation}
under the following tetrads $e^{i}_{\alpha}$ transformations;
\begin{equation}
        (e^{i}_{\alpha})'= exp(\rho) e^{i}_{\alpha}.
\label{cor1}
\end{equation}
The spin connections $\omega^{i}_{j\alpha}$ are conformally invariant like
other gauge fields.

The affine connections and the torsions which are the antisymmetric
components of the affine connections transform as follows;
\begin{equation}
    (\Gamma^{\gamma}_{~\beta \alpha})'
    =\Gamma^{\gamma}_{~\beta \alpha} + \delta^{\gamma}_{~\beta} \partial_{\alpha}
    \rho, ~~~
    (T^{\gamma}_{~\beta \alpha})'=
    T^{\gamma}_{~\beta \alpha}+
    \delta^{\gamma}_{\beta}\partial_{\alpha}\rho
    - \delta^{\gamma}_{\alpha}\partial_{\beta}\rho.
\label{tor1}
\end{equation}
Therefore, the trace of the torsion $T^{\gamma}_{~\gamma \alpha}$
effectively plays the role of a conformal gauge field.
In general, the torsion can be decomposed into three components;
\begin{equation}
T^{\alpha}_{~\beta\gamma}=
        \Sigma^{\alpha}_{~\beta \gamma}
         + A^{\alpha}_{~\beta \gamma}
         - \delta^{\alpha}_{\gamma}S_{\beta}
         + \delta^{\alpha}_{\beta}S_{\gamma},
\label{tor2}
\end{equation}
where $\Sigma_{[\alpha\beta\gamma]} \equiv 0, ~~
\Sigma^{\alpha}_{~\alpha\gamma} \equiv 0, ~~
A_{\alpha\beta\gamma} \equiv T_{[\alpha\beta \gamma]}$.
The traceless part of torsion $C^{\alpha}_{~\beta \gamma} \equiv
\Sigma^{\alpha}_{~\beta\gamma} + A^{\alpha}_{~\beta\gamma}$ is
conformally invariant.
\begin{equation}
          (S_{\alpha})' = S_{\alpha} +
          \partial_{\alpha}\rho, ~~~
          (C^{\alpha}_{~\beta \gamma})'=
          C^{\alpha}_{~\beta \gamma} .
\label{vec}
\end{equation}

Because the minimal extension to Riemann-Cartan space is sufficient for
our purpose, we impose the conformally invariant torsionless
condition;
\begin{equation}
          C^{\alpha}_{~\beta \gamma} \equiv 0.
\label{zero}
\end{equation}
This condition is the conformally invariant extension of the
torsionless condition in Riemann space,
$ T^{\alpha}_{~\beta \gamma}\equiv 0$.
For this minimally extended Riemann-Cartan space, the affine connection can
be written in terms of
$g_{\mu\nu}$ and $S_{\alpha}$;
\begin{equation}
          \Gamma^{\alpha}_{~\beta\gamma}=
          \{^{\alpha}_{\beta\gamma}\}+
          S^{\alpha} g_{\beta\gamma} -
          S_{\beta}\delta^{\alpha}_{\gamma}.
\label{cristo}
\end{equation}

Introducing the conformally covariant derivative $D_{\alpha}$ for scalar
field $\phi$,
\begin{equation}
D_{\alpha}\phi \equiv \partial_{\alpha}\phi + S_{\alpha}\phi,
\end{equation}
we have an extended conformal coupling of induced gravity
up to total derivatives as follow;
\begin{equation}
      I =\int d^{4}x \sqrt{-g}
        \{\frac{\xi}{2}R(\Gamma)\phi^{2} +
        \frac{1}{2}D_{\alpha}\phi D^{\alpha}\phi -
        \frac{1}{4}H_{\alpha\beta}H^{\alpha\beta} -
        V(\phi)\},
\label{gen}
\end{equation}
where we have excluded the curvature square terms.
Now, the coupling $\xi$ is a dimensionless arbitrary constant.
Using Eq.(\ref{cristo}) we can rewrite this action in
terms of Riemann curvature scalar $R(\{\})$;
\[
      I=\int d^{4}x \sqrt{-g}
      \{\frac{\xi}{2}R(\{\})\phi^{2}
        + \frac{1}{2}\partial_{\alpha}\phi\partial^{\alpha}\phi
        + (1-6\xi) S^{\alpha}(\partial_{\alpha}\phi)\phi
\]
\begin{equation}
        + \frac{1}{2}(1-6\xi)S_{\alpha}S^{\alpha}\phi^{2}
        - \frac{1}{4}H_{\alpha \beta}H^{\alpha \beta}
        - V(\phi)\}.
\label{iaction}
\end{equation}
The more general form of induced gravity action can be considered \cite{buch1,buch2},
but we restrict the couplings to be conformal.
In the limit of $\xi \rightarrow \frac{1}{6}$, this extended conformal
coupling is reduced to the ordinary conformal coupling in Riemann space
decoupled from the vector torsion.

\section{Dilaton Gravity from Induced Gravity with Gradient Torsion}

Analyzing the equations of motion for the action Eq.(\ref{iaction}) with an
effective potential $V_{eff}(\phi;S_{\alpha},g_{\beta\gamma})$ which depends
on metric and torsion in general, we obtain the following two equations of
motion and a constraint for the scalar potential;
\begin{equation}
\nabla_{\mu}H^{\mu\nu}=
-(1-6\xi)\{(\partial^{\nu}\phi)\phi+S^{\nu}\phi^{2}\}
+\frac{\partial V_{eff}(\phi;S_{\alpha},g_{\beta\gamma})}{\partial S_{\nu}}~,
\label{box1}
\end{equation}
\[
\xi\phi^{2}G_{\mu\nu}=
  (H_{\mu\alpha}H_{\nu}^{~\alpha}
   -\frac{1}{4}g_{\mu\nu}H_{\alpha\beta}H^{\alpha\beta})
- (\partial_{\mu}\phi\partial_{\nu}\phi
  -\frac{1}{2} g_{\mu\nu}\partial_{\alpha}\phi\partial^{\alpha}\phi)
- (1-6\xi)\phi^{2}(S_{\mu}S_{\nu}-\frac{1}{2}g_{\mu\nu}S_{\alpha}S^{\alpha})
\]
\[
- (1-6\xi)(S_{\mu}\phi\partial_{\nu}\phi+S_{\nu}\phi\partial_{\mu}\phi
   - g_{\mu\nu}S^{\alpha}\phi\partial_{\alpha}\phi)
+ \xi\{\nabla_{\mu}(\phi\partial_{\nu}\phi)+\nabla_{\nu}(\phi\partial_{\mu}\phi)
       -g_{\mu\nu}\Box\phi^{2}\}
\]
\begin{equation}
- g_{\mu\nu}V_{eff}(\phi;S_{\alpha},g_{\beta\gamma})
+ 2\frac{\partial V_{eff}(\phi;S_{\alpha},g_{\beta\gamma})}
        {\partial g^{\mu\nu}}~,
\label{long}
\end{equation}
\begin{equation}
4V_{eff}(\phi;S_{\alpha},g_{\beta\gamma})
-\phi\frac{\partial V_{eff}(\phi;S_{\alpha},g_{\beta\gamma})}{\partial\phi}
=2\frac{\partial V_{eff}(\phi;S_{\alpha},g_{\beta\gamma})}
{\partial g^{\mu\nu}}g^{\mu\nu}
+\nabla_{\nu}\frac{\partial V_{eff}(\phi;S_{\alpha},g_{\beta\gamma})}
{\partial S_{\nu}}~,
\label{combix}
\end{equation}
where all covariant derivatives are in Riemann space with the
Christoffel connections \cite{yoon95b,yoon97}.

The constraint Eq.(\ref{combix}) requires that the metric independent bare
potential should be quartic in the scalar field,
$V_{o}(\phi)=\frac{\lambda}{4!}\phi^{4}$, and
the deviation of the radiatively corrected effective potential from the
quartic form is only allowed with the compensation by the metric and vector
torsion dependencies of the effective potential.
Because this constraint comes from the assumption that
the bare action is conformally invariant except the potential term,
if we consider non-conformal coupling in kinetic and interacting terms,
such a constraint would not appear.

Let us consider a reduction of the system. If the effective potential does not have
the vector torsion dependency, i.e. $V_{eff}(\phi;g_{\beta\gamma})$, Eq.(\ref{box1})
allows the following conformally invariant reduction;
\begin{equation}
D_{\alpha}\phi = 0.
\end{equation}
This implies that the vector torsion is a gradient form;
\begin{equation}
S_{\alpha}=-\partial_{\alpha}ln(\phi/\phi_{o}) =  -
\partial_{\alpha}\sigma, ~~~ \phi \equiv \phi_{o}e^{\sigma},
\end{equation}
where $\phi_{o}$ is a dimensional constant,
and the field strength of the vector torsion vanishes
$H_{\alpha\beta}=0$, which is consistent with Eq.(\ref{box1}).
In this reduction, the bare action of Eq.(\ref{gen}) becomes
\begin{equation}
      I =\int d^{4}x \sqrt{-g}\phi_{o}^{2}e^{2\sigma}
        \{ \frac{\xi}{2}R(\{\})
        + 3\xi\partial_{\alpha}\sigma \partial^{\alpha}\sigma
        - \frac{\lambda}{4!}\phi_{o}^{2}e^{2\sigma} \}.
\label{gen1}
\end{equation}
This is the form of conformal factor theory of dilaton gravity \cite{odintsov1,odintsov2}.
If the antisymmetric torsion
term is $\frac{1}{12}C_{\alpha\beta\gamma}C^{\alpha\beta\gamma}$ included,
this action is the form of string gravity with some redefinition of fields
except the quartic potential term \cite{Saa}.

For the reduction of the effective action, the potential term is replaced by
$e^{-2\sigma}V_{eff}(\phi_{o}e^{\sigma},g_{\alpha\beta})$. In this
reduction, a dimensional constant $\phi_{o}$ is introduced, but
the action is invariant under the global scaling $dx^{\mu} \rightarrow a
dx^{\mu}$, $\phi_{o} \rightarrow \phi_{o}/a$ if no
conformal anomaly is introduced in the effective action. However,
the appearance of the conformal anomaly in the conformally
induced gravity is not allowed due to the constraint
Eq.(\ref{combix}) which is the requirement of conformal invariance
\cite{yoon97} from the consistency of equations of motion for the bare
and the effective action in the conformally induced gravity \cite{yoon95b}.
In this reduction, the constraint which effective potential should satisfy is
\begin{equation}
4V_{eff}(\phi;g_{\beta\gamma})
-\phi\frac{\partial V_{eff}(\phi;g_{\beta\gamma})}{\partial\phi}
=2\frac{\partial V_{eff}(\phi;g_{\beta\gamma})}
{\partial g^{\mu\nu}}g^{\mu\nu}~.
\label{combix0}
\end{equation}
Therefore, only metric independent quartic potential with effective
coupling $\lambda_{eff}$ seems to be allowed. But, in general, the
quantum effect by quantum fluctuation of scalar field on
the classical metric and torsion background \cite{odintsov1,odintsov2} gives
the following corrections to the scalar mass $m^{2}$, the coupling
$\lambda$ and the cosmological constant $\Lambda$ respectively \cite{Coleman,Dewitt}.
\begin{equation}
\delta m^{2} \propto \lambda m^{2},~~~
\delta\lambda \propto \lambda^{2},~~~
\delta\Lambda \propto a m^{2} + b \lambda.
\end{equation}
Therefore, $\lambda=0$ would be the only solution of
Eq.(\ref{combix0}) as a trivial fixed point of the renormalization group.
However, the definite claim about $\lambda=0$ is possible only after the
consideration of full quantum effects of the theory, which is of course
far beyond of our scope yet.

Redefining the metric,
\begin{equation}
g_{\alpha\beta}e^{2\sigma} \rightarrow g_{\alpha\beta},
\end{equation}
the above dilaton gravity action can be written in the following
standard Einstein action;
\begin{equation}
      I =\int d^{4}x \sqrt{-g}
        \{ \frac{\xi}{2}\phi_{o}^{2}R(\{\})
        - \frac{\lambda}{4!}\phi_{o}^{4} \},
\label{gen2}
\end{equation}
where the gravitational constant $G_{N}=
\frac{1}{8\pi\xi\phi_{o}^{2}}$, and the cosmological constant
$\Lambda = \frac{\lambda}{4!}(\frac{1}{8\pi\xi G_{N}})^{2}$.
If the effective potential deviates from the quartic form, then
there would remain explicit $\sigma$ dependency in the action
after this redefinition of metric even though it is not plausible due to the
constraint Eq.(\ref{combix0}).

In this Einstein frame the conformal symmetry is hidden and the
cosmological constant term has the origin of the quartic potential
term in the original induced gravity action.
Reminding the constraint Eq.(\ref{combix0}) which the effective
potential should satisfy, the non-zero coupling $\lambda$ can be
hardly expected after the radiative correction by the scalar field
in the original induced gravity action \cite{Coleman,Dewitt}.
Therefore, we can say that if the Einstein gravity
have the root in the conformally induced gravity, the non-zero cosmological
constant is not plausible due to the constraint from the conformal
coupling.

In this discussion, we have not considered torsions generated by
matter fields because vector torsion couplings are not expected in the standard
minimal action for Dirac and gauge fields \cite{buch2}.

~\\~\noindent
{\it Acknowledgments:}
This work was supported in part by Korean Ministry of Education through
the program BSRI-2441.
The author wishes to acknowledge the financial support of Hanyang University,
Korea, made in the program year of 1998.

\end{document}